\documentclass[10pt,leqno]{amsart}
\usepackage{graphicx}
\usepackage{indentfirst}

\usepackage{amssymb,amsthm,amsmath}
\usepackage{xcolor,paralist,titlesec,fancyhdr,etoolbox}
\makeatletter
\patchcmd{\@settitle}{\uppercasenonmath\@title}{}{}{}
\newcommand{\paperaffiliations}{%
  \par\vspace{0.75em}%
  \textsuperscript{1}Nanyang Technological University\par
  \textsuperscript{2}Xiamen University\par
  \textsuperscript{3}Shanghai Jiaotong University\par
  \textsuperscript{4}Shanghai University of Finance and Economics%
}
\patchcmd{\@setauthors}{\MakeUppercase{\authors}}{\authors\paperaffiliations}{}{}
\makeatother
\usepackage{fontspec}
\usepackage{listings}
\definecolor{leanblue}{RGB}{0,70,150}
\definecolor{leangreen}{RGB}{0,110,70}
\definecolor{leanpurple}{RGB}{110,35,130}
\lstdefinelanguage{Lean}{
  keywords={abbrev,answer,by,class,def,else,example,fun,if,import,in,inductive,
    instance,let,namespace,open,Prop,structure,then,theorem,Type,where},
  keywordstyle=\color{leanblue}\bfseries,
  morekeywords=[2]{DecidableEq,Finset,Fintype,IsAllocation,IsEFX,Nonempty,Pref,
    Reflexive,SocialChoice,Transitive},
  keywordstyle=[2]\color{leanpurple},
  morecomment=[l]{--},
  morecomment=[s]{/-}{-/},
  commentstyle=\color{leangreen}\itshape,
  morestring=[b]",
  stringstyle=\color{leangreen},
  sensitive=true,
  alsoletter={_'}
}
\lstnewenvironment{leancode}
  {\lstset{language=Lean,basicstyle=\ttfamily\footnotesize,breaklines=true,
    columns=fullflexible,keepspaces=true,showstringspaces=false,upquote=true,
    frame=single}}
  {}
\newcommand{\leaninline}[1]{\lstinline[language=Lean,basicstyle=\ttfamily\small,columns=fullflexible]!#1!}
\usepackage{csquotes,hyperref}
\titleformat{\section}{\normalfont\Large\bfseries}{\thesection.}{0.5em}{}
\titleformat{\subsection}{\normalfont\large\bfseries}{\thesubsection.}{0.5em}{}
\titlespacing*{\section}{0pt}{2ex}{1ex}
\titlespacing*{\subsection}{0pt}{1.5ex}{0.75ex}

\hypersetup{ colorlinks=true, linkcolor=black, filecolor=black, urlcolor=black }

\begin{document}
\title{EconCSLib: A Lean Library for Computational Economics \\ and AI-Assisted Research} 
\author{Xiaohui Bei\textsuperscript{1}}
\author{Jiajun Ma\textsuperscript{2}}
\author{Zhan Jing\textsuperscript{3}}
\author{Hongfei Fu\textsuperscript{3}}
\author{Zhihao Gavin Tang\textsuperscript{4}}
\date{}

\begin{abstract}
Mathematical formalization uses interactive theorem provers to turn 
informal mathematical statements into machine-checkable artifacts.  The success of mathlib,
a large collaborative library for Lean, illustrates the potential of this
approach.
Recent progress in AI-assisted programming and theorem proving is also making
large-scale formalization more practical.  This paper presents EconCSLib, an
early Lean 4 library for computational economics, as both infrastructure and a case study for AI-assisted formalization.  
The library aims to provide reusable definitions and theorems for game theory, 
mechanism design, social choice, and related areas.  
Beyond verified proofs of existing results, the library also aims to host machine-checked open problems and formalization of modern research papers.  
We discuss the design principles behind the library, the lessons learned from its development, and future directions for AI-assisted formalization in computational economics.
\end{abstract} 

\maketitle

\bigskip

\section{Introduction}

Mathematical formalization turns definitions, theorems, and proofs into objects
that can be checked by a computer. Interactive theorem provers such as Lean \cite{lean}, Rocq \cite{RocqProver}, Isabelle \cite{Nipkow2002-NIPIAP} make
this possible at scale: they construct a formal language for mathematics by providing a proof
checker and an environment for building reusable libraries of verified results.
At the heart of the Lean 4 ecosystem is mathlib \cite{mathlib}, a large library of formalized mathematics that has been developed by a community of volunteers. 
The development of mathlib has shown that large parts of modern mathematics can be formalized in a way that is both rigorous and reusable. This new infrastructure has enabled researchers to build on each other's work more easily and to explore new connections between different areas of mathematics, and has gained significant traction in the mathematical community \cite{taoMachineAssistedProof2025,kontorovich2025}.
The success of mathlib has also inspired efforts to build similar libraries for other domains, such as cslib for computer science \cite{barrettCSLibLeanComputer2026}, physlib for physics \cite{tooby-smithHepLeanDigitalisingHigh2025b}, chemistry-physics \cite{bobbinFormalizingChemicalPhysics2024}, and cryptography \cite{tumaVCVioVerifiedCryptography2026}.
The goal of these libraries is to provide a foundation for formalizing results in their respective fields and to enable researchers to build on each other's work in a more systematic way.

Computational economics, also known as EconCS, is a natural target for this style of formalization.
The field combines mathematical modeling and
algorithmic design, as reflected in standard treatments of algorithmic game
theory and mechanism design \cite{agt}. Its basic objects---players,
preferences, strategies, mechanisms, equilibria, allocations, and solution
concepts---are highly structured, and small changes in assumptions can alter
the validity of a theorem. 
Formalization of these results can therefore provide a rigorous foundation for the field.
There has been some prior work on formalizing specific topics and results in computational economics, such as basic game theory \cite{lean-gametheory}, combinatorial games \cite{lean-combinatorial-games}, voting theory in social choice \cite{hollidayVotingTheoryLean2021c,lean-socialchoice}, and mechanism design \cite{caminati2015,jouvelot2022foundational}.
Recent work in fair division has also used formalization to verify the correctness of results \cite{akrami2026}.
However, there has not yet been a systematic effort to build a comprehensive library of formalized definitions and theorems for computational economics. Such
a shared library of definitions and theorems can make it easier for researchers to build on each other's work and to explore new connections between different areas of computational economics.

Recent progress in formalization has also been driven by advances in AI-assisted programming and theorem proving. Large language models (LLMs) and other AI tools can now generate formal proofs \cite{deepseekProverV2,aristotle} and translate informal mathematics into formal statements \cite{gauss}. This has greatly reduced the proof-engineering cost
that has historically limited formalization. These tools do not remove the need for human
mathematical judgment, especially in choosing definitions and verifying that a
formal statement matches the intended theorem. They do, however, make it more
plausible that domain researchers can participate directly in formalization
without becoming full-time proof engineers.

\subsection{Contributions}
In this work, we present EconCSLib\footnote{The current EconCSLib codebase is at \url{https://github.com/gametheoryinlean/EconCSLib}.}, an early infrastructure effort for
AI-assisted formalization in computational economics. 
EconCSLib is a Lean 4 library that provides reusable definitions, theorems, and algorithms for a range of topics in EconCS.  Its goal is not only to verify isolated results, but to build shared formal infrastructure that can be extended and reused across different areas of computational economics.

EconCSLib develops common Lean
interfaces for foundamental concepts in EconCS such as players, preferences, profiles, utilities, allocations,
mechanisms, and related mathematical objects.  These interfaces are used
across several branches of the library, including game theory, mechanism design,
social choice, and optimization-related
components.  The current snapshot contains formalized definitions and results
ranging from classical theorems to executable algorithms and checkers, with the
aim of making later formalizations build on a common vocabulary rather than
recreating local models for each theorem.

The design philosophy of the library is organizing EconCS concepts around small
core abstractions, introducing additional assumptions only where they are
needed, and connecting theorem-oriented formalization with executable
content. This design is especially important in EconCS, where the same objects
often appear in different forms across different subfields, 
and where small changes in assumptions can change the truth of a statement.

Another contribution of the library is to treat open problems and future formalization
targets as part of the library.  EconCSLib is designed to host machine-checked
open problems and partially formalized research programs.  
This makes it possible to record precise formal statements of these open
questions, accumulate partial results against shared definitions, and provide
benchmarks for AI-assisted theorem-proving systems.

Finally, we use the development of EconCSLib as a case study in AI-assisted
formalization.  We discuss where current AI tools are helpful, where human
mathematical judgment remains essential, and how an interactive workflow between
domain researchers, AI assistants, and the Lean kernel can reduce the cost of
building reusable formal infrastructure.  We also outline future directions for
the project and its potential role in computational economics research.

\subsection{Concurrent Work}

Independently and contemporaneously with this project, Garg introduced a Lean 4
library and workflow with the same name, EconCSLib, for AI-assisted
formalization in economics and computation \cite{gargEconCSLib2026}.  The two
projects share a broad vision: both aim to make formalization and Lean useful for EconCS research,
both treat reusable domain infrastructure as an important outcome of
formalization, and both rely on a human-AI-Lean workflow in which AI tools 
help produce formalization code while humans remain responsible for validating the
translation from mathematical intent to formal statements.

The emphasis of the two projects is different, however. Garg's EconCSLib is organized
primarily around formalizing research papers: each paper has paper-facing Lean
interfaces, dependency DAGs, validation reports, and review dashboards, with
reusable lemmas promoted into shared infrastructure when they arise from the
paper formalizations.  Our EconCSLib is organized primarily as reusable domain
infrastructure for computational economics: it develops common abstractions for
foundational concepts in EconCS and related mathematical interfaces, and it is 
intended to host not only verified classical results but also formally stated open problems.  
These differences make the two projects complementary rather than redundant.
Garg's primary goal is to help the EconCS community formalize its research
papers, and reusable infrastructure is one of the main bottlenecks for such a
workflow.  To the extent that our project supplies stable definitions,
interfaces, and general-purpose lemmas, it can make paper-by-paper formalization
substantially easier.  Conversely, a paper-oriented workflow can identify which
abstractions are most useful in practice and can feed reusable components and
test cases back into the infrastructure-first library.

\section{EconCSLib: Architecture and Current Progress}

EconCSLib is a Lean 4 library for computational economics, built on mathlib and
intended to serve as reusable research infrastructure rather than as a set of
isolated formal proofs. The long-term vision is that a researcher should be
able to import a common vocabulary for various concepts in EconCS 
and then use this vocabulary to verify their results or to prove new theorems. 
In this sense, the library is meant to
play for computational economics a role analogous to the one mathlib plays for
mathematics.

\subsection{Design Principles}
The first design principle of EconCSLib is reuse through abstraction. The library avoids
building a separate local model for each theorem. Instead, it organizes the
field around common semantic objects such as profiles, preference relations,
and allocation rules, so that a definition introduced for one area can be reused in another. For example,
\textit{preferences} appear in mechanism design, social choice, and matching; 
the concept \textit{strategy profiles} is shared in both game theory and mechanism design.

\begin{figure}[tbp]
\begin{leancode}
class IsPreference {A : Type*} (R : A → A → Prop) : Prop where
  reflexive : Reflexive R
  transitive : Transitive R
  total : ∀ a b : A, R a b ∨ R b a

structure Pref (A : Type*) where
  rel : A → A → Prop
  prop : IsPreference rel

def PrefProfile (N A : Type*) := N → Pref A

structure SocialChoice.Instance (N A : Type*) where
  feasible : A → Prop
  pref : N → Pref A
\end{leancode}

\caption{Reusable preference and social-choice primitives. \leaninline{IsPreference}
records reflexivity, transitivity, and completeness. The bundled relation is
shared by preference profiles and generic social-choice instances.}
\label{fig:code-social-choice-primitives}
\end{figure}

The second principle is to keep assumptions local. Core definitions should
contain only the data needed to define the object itself. 
Additional assumptions and structure are added at theorem sites when they are actually needed. This is
especially important in economics, where small changes in modeling assumptions
can change the content of a theorem. For instance, in game theory and mechanism design, 
profiles and unilateral deviations can be defined without finiteness assumptions, while existence theorems for equilibria or algorithms for computing them can introduce the necessary finiteness and decidability assumptions locally. 

Figure \ref{fig:code-social-choice-primitives} and \ref{fig:code-strategic-game-foundation} are Lean code snippets illustrating these principles. The first shows a preference relation defined as a bundled structure with an associated class for the standard properties of a preference relation. The second shows the core definitions for strategic games, including the strategy spaces and payoff functions, as well as the definition of a profile and a unilateral deviation operation.


\begin{figure}[tbp]
\begin{leancode}
structure StrategicGame (N : Type*) (U : Type*) where
  strategy : N → Type*
  payoff : ((i : N) → strategy i) → N → U

namespace StrategicGame

abbrev Profile (G : StrategicGame N U) :=
  (i : N) → G.strategy i

abbrev deviate {G : StrategicGame N U} [DecidableEq N]
    (profile : G.Profile) (i : N) (s' : G.strategy i) :
    G.Profile :=
  Function.update profile i s'
\end{leancode}
\caption{Core game-theoretic abstractions. A strategic game stores only strategy
spaces and payoffs; game-bound profiles do not require finiteness assumptions, while the deviation operation is defined for any profile and any player, with no assumptions on the strategy spaces. 
These definitions are intended to be reused across different game-theoretic developments, and
provide a shared interface for later definitions.}
\label{fig:code-strategic-game-foundation}
\end{figure}


\subsection{Architecture and Coverage}

\begin{figure}[t]
\centering
\includegraphics[width=\textwidth,height=.78\textheight,keepaspectratio]{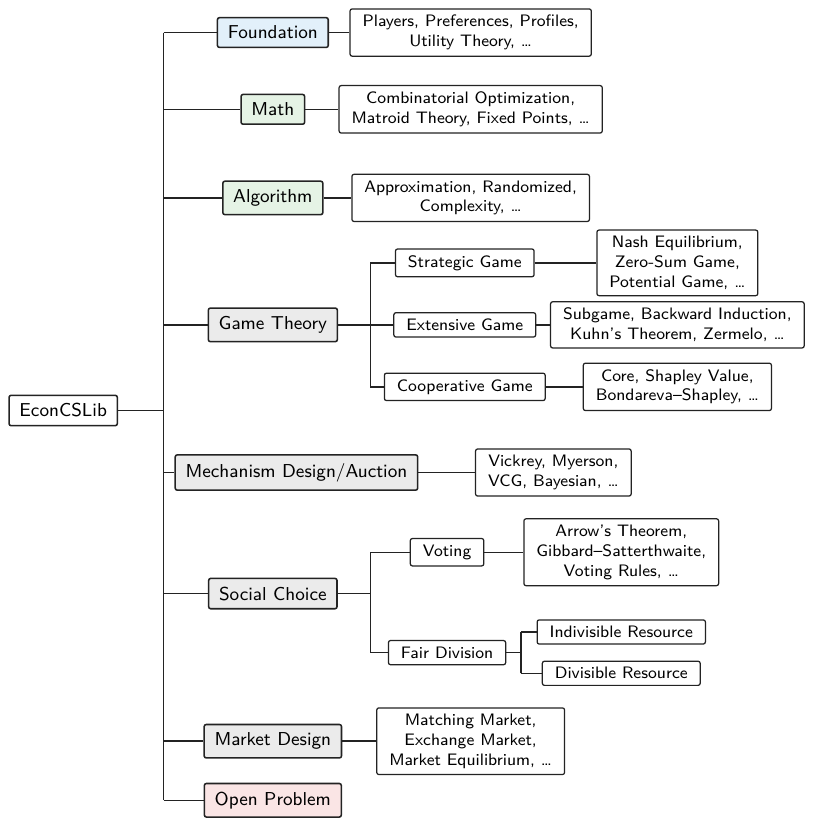}
\caption{A high-level organization of EconCSLib.}
\label{fig:econcslib-organization}
\end{figure}

Figure~\ref{fig:econcslib-organization} summarizes the present organization of
the library. Near the top are foundation modules for players, preferences, profiles, and utility theory. 
These modules provide the common vocabulary that is reused across the rest of the library:
profiles occur in games and mechanisms, preferences occur in voting, fair division, and matching, 
and utility representations connect ordinal
statements with numerical payoff models when a theorem requires them.
These modules are kept small and abstract, so that they can be reused in different contexts without forcing a particular modeling choice.

Besides the foundation modules, the library also contains
two additional infrastructure branches that support the domain-specific developments.
\begin{itemize}
    \item The mathematical branch collects results and interfaces from
combinatorial optimization, matroid theory, fixed-point arguments, and related
finite-dimensional tools. 
These mathematical tools play a supporting role in many theorems across the field.
    \item The algorithmic branch contains the assumptions and
objects needed for executable developments and for formalizing algorithmic results.
These include: approximation algorithms, randomized algorithms, online algorithms, 
complexity-theoretic statements, decidable predicates, and computable checkers. 
\end{itemize} 

The lower branches of the figure show the main EconCS domains currently
represented in the library. The game-theory branch includes strategic-form,
extensive-form, and cooperative games. Strategic games support standard
solution concepts such as dominance, best response, mixed strategies, Nash equilibrium, 
zero-sum and constant-sum structure, and potential games.
Extensive games provide game trees, subgames, backward induction, and a
machine-checked Zermelo-style determinacy result, with targets such as Kuhn's
theorem. Cooperative-game
infrastructure includes concepts such as the core and Shapley value, together
with longer-term targets such as Bondareva--Shapley.

The mechanism-design branch covers both dominant-strategy mechanisms, Bayesian mechanisms, 
and auctions under the incomplete-information game paradigm. It
includes verified results for Vickrey auction, Myerson's optimal auction and VCG-style
mechanisms.

The social-choice branch divides into voting and fair division. Voting modules
contain voting rules, axioms, and machine-checked proofs of Arrow's theorem and
Gibbard--Satterthwaite. Fair division modules distinguish
indivisible and divisible resources, and cover a range of fairness notions such as 
envy-freeness, proportionality, and their relaxations.

The market-design branch covers matching markets, exchange markets, and market
equilibrium formalizations. It includes a verified formalization of stable
matching and Gale--Shapley's deferred acceptance algorithm, with targets such as
competitive equilibrium with equal incomes and the existence of market
equilibria under various assumptions.

\medskip

EconCSLib is still an early-stage library, and the coverage of the branches is uneven.
Different branches currently contain different kinds of content:
some have complete proofs, others primarily have definitions and formally stated targets for future work.
We expect the coverage to evolve over time, with more theorems and algorithms added as the library matures. 
Nevertheless, the library already contains a substantial collection of verified
proofs. Representative completed formalizations include Arrow's theorem,
Gibbard--Satterthwaite, Zermelo-style determinacy, linear-programming strong
duality and Farkas' lemma, minimax via Loomis' theorem, the Brouwer, KKM, and
Scarf fixed-point theorems, VCG mechanisms, Myerson's optimal auction,
Gale--Shapley's deferred acceptance algorithm, the cut-and-choose procedure and dubins-spanier procedure for cake cutting, and the existence of EFX allocations for two agents. At the snapshot described in this paper, EconCSLib
contains over 40,000 lines of Lean code and over 1,300 theorems/lemmas, without any additional axioms.


\subsection{Open Problems as First-Class Objects}

EconCSLib is intended to host not only verified results but also formally stated
open problems.  An open problem can be represented by a \leaninline{sorry}-bearing
declaration or by a clearly delimited proposition whose proof remains to be
supplied.  Even before the problem is solved, Lean checks that the statement is
well typed, that it uses the shared definitional framework, and that its
assumptions are explicit.

Treating open problems as first-class library objects has several benefits:
\begin{itemize}
\item open problems become precise, machine-checkable mathematical artifacts
rather than informal English descriptions whose interpretation may vary;
\item maintaining a shared list gives researchers a common reference point for
recording partial results and special cases and avoiding redundant efforts;
\item autonomous AI problem-solving tools can target the formal statements
directly, with proposed proofs or disproofs checked by the Lean kernel;
\end{itemize}

This approach is similar to the Formal Conjectures project
\cite{firschingFormalConjecturesOpen2026}, but specialized to computational
economics.  We also plan to adopt that project's \leaninline{answer(sorry)}
mechanism.  This mechanism distinguishes a missing mathematical answer from a
missing proof: for example, a solver may need to replace a placeholder with a
number, a function, or a truth value and then prove that the proposed answer is
correct.  It therefore supports open problems that are not naturally expressed
as a theorem with an already fixed conclusion.

\begin{figure}[tbp]
\begin{leancode}
theorem existsDemandOracleSWMAlgorithmBeatingOneMinusInvE
    (I Omega : Type*) {alpha : Type*} [Fintype I] [Fintype Omega] [DecidableEq alpha]
    (M : Finset alpha) [Fintype (BundlePartitionAllocation I M)]
    [Nonempty (BundlePartitionAllocation I M)] :
    answer(sorry) ↔
      ∃ eps : ℝ, 0 < eps ∧
        ∃ alg : RandomizedSubmodularWelfareAlgorithm I Omega M,
          alg.IsPolynomial ∧
            ∀ profile : SubmodularWelfareMaximizationProfile I M,
              (1 - 1 / Real.exp 1 + eps) *
                  OPT (fun i => (profile.oracle i).valuation) ≤
                expectedBundlePartitionSocialWelfare alg profile := by
  sorry
\end{leancode}
\caption{Open problem statement for submodular welfare maximization with demand
oracles.}
\label{fig:code-open-swm}
\end{figure}

\begin{figure}[tbp]
\begin{leancode}
theorem existsEFXAllocationForAdditiveValuations :
    answer(sorry) ↔
      ∀ (N G : Type*) [Fintype N] [Nonempty N] [DecidableEq G],
        ∀ (w : AdditiveValuation N G)
          (hnn : ∀ i g, 0 ≤ w.weight i g) (allGoods : Finset G),
          ∃ A : Allocation N G,
            IsAllocation allGoods A ∧ IsEFX w.toValuation A := by
  sorry
\end{leancode}
\caption{Open problem statement for existence of EFX allocations of indivisible goods
with additive nonnegative valuations.}
\label{fig:code-open-efx}
\end{figure}

Figures~\ref{fig:code-open-swm} and~\ref{fig:code-open-efx} illustrate two example open problems.  
The first asks for a randomized polynomial-time demand-oracle algorithm for submodular welfare
maximization whose approximation guarantee beats the \(1-1/e\) barrier.  The
formal statement is built on library interfaces for submodular valuations,
demand oracles, randomized algorithms, and expected welfare.

The second asks whether complete EFX allocations always exist for indivisible
goods with additive nonnegative valuations. This is one of the most prominent open
problems in fair division. Existence is known for up to three
agents \cite{chaudhury2024}; the case of four or more agents remains open.  Results for small numbers
of agents can be formalized against the same definitions and reused as
incremental progress toward the general problem.

\section{Lessons from AI-Assisted Formalization and Outlook}

The initial development of EconCSLib benefited significantly from AI-assisted programming and theorem proving tools. 
But the value of AI assistance is unevenly distributed across the workflow.
We find current AI tools to be very effective at routine proof engineering: expanding
definitions, filling short proofs, and translating an informal argument into an
initial Lean statement.  They are also useful for iterative maintenance tasks,
such as adapting proofs after a definition is revised or a module is
reorganized.  These capabilities greatly reduce the cost of experimentation. 

The harder tasks are architectural and mathematical rather than syntactic.
An AI assistant may specify statements and extend modules that do not capture the intended mathematical object or that is not reusable for later work.  
This makes human oversight particularly important at the level of supervising 
the semantic mathematical objects.  The most effective workflow has
been interactive:  A domain researcher specifies the intended mathematical
object and checks whether the formal statement captures the relevant economic
model; AI tools accelerate the local construction and repair of definitions and
proofs; and the Lean kernel verifies the resulting artifact.

These lessons suggest that building a reusable library is significantly more demanding than formalizing a
single isolated result.  Each design choice in the library's definitions and theorems
becomes an input to future proofs and algorithms.  EconCSLib's
emphasis on small core objects and local assumptions is partly a response to
this observation.  The library should make the general object easy to state and
allow additional structure to be introduced only where it is used.

\subsection{Outlook}
Looking forward, the immediate next step for EconCSLib is to extend the library from its current foundation
into a broader collection of verified results and reusable domain interfaces.
In addition, we identify several longer-term directions for this project and for AI-assisted formalization in computational economics more generally.

\begin{compactitem}
\item \emph{Formalization of contemporary results.}  We hope to formalize a selection of modern research papers in computational economics, starting with results that are already well understood and then moving toward more recent and less familiar work.
Formalizing such modern EconCS papers would test whether the current abstractions are expressive enough for
research-level arguments. It would also reveal missing mathematical infrastructure in the library, 
and produce reusable components for later work.
\item \emph{AI-assisted theorem discovery and counterexample search.}  
Formal definitions provide a precise search space in which candidate 
solutions and counterexamples can be generated automatically by AI tools
and then checked by the Lean kernel. With the rapid growth of AI capabilities, 
we expect this to become an increasingly important part of the research workflow.
As a result, the library's collection of open problems and partially formalized research programs will become a valuable resource for testing and evaluating AI problem-solving tools in computational economics.
\item \emph{Benchmarks for theorem-proving agents.}  EconCSLib can supply
domain-specific benchmarks ranging from local proof obligations to open
research problems. Such benchmarks are essential for evaluating the performance of AI-assisted theorem-proving agents and for guiding their development toward the needs of computational economics research.
\end{compactitem}

These directions reinforce each other.  Formalizing new papers improves the
library's abstractions; the resulting reusable components make it easier to formalize subsequent papers; and the open problems and benchmarks provide a testbed for AI tools that can further accelerate the formalization of new results.
The long-term role of EconCSLib is therefore not merely
to archive finished proofs.  It is intended to provide shared,
machine-checkable infrastructure for future AI-assisted research in
computational economics.

\subsection{Open-Source and Community Development}

We plan to develop EconCSLib as an open-source, community-driven project.  After
the initial release, we hope that much of the library's new content will be
contributed by researchers interested in computational economics and formal
methods.  Mathlib \cite{mathlib} and cslib
\cite{barrettCSLibLeanComputer2026} provide useful guidelines: both emphasize
shared definitions, reusable results, public review, and sustained collaboration
rather than a collection of isolated formalization projects.

EconCSLib also has some domain-specific advantages.  Computational economics is
more narrowly scoped than mathematics or computer science as a whole, which may
allow the community to converge more quickly on common abstractions and review
new contributions more rapidly.  We also expect AI tools to play a more substantial
role in contributing to the library, both by generating new content and by assisting with the review of proposed 
contributions. Human review will remain important for
checking mathematical meaning and maintaining coherent interfaces.

Following mathlib and cslib, we plan to establish an EconCSLib channel on the
Lean Zulip chat.  This channel would provide a forum for active and open discussing 
among everyone interested in the library.  Our goal is to foster an active and open
community around the library's continued development.

\section{Conclusion}

EconCSLib is an initial step toward a shared formal foundation for
computational economics.  Its current development shows that concepts and results
from across the field can be organized
within a common Lean library.  AI tools are making this effort substantially more
practical, while the design of reusable abstractions and the interpretation of
formal statements remain important tasks for domain experts.
The long-term goal is not simply to verify a collection of existing results,
but to build infrastructure for future research.  As EconCSLib grows through
open-source collaboration, we hope that it will support more reliable
formalization, computation, and AI-assisted discovery in computational
economics.

\bibliographystyle{amsplain}
\bibliography{references}
\end{document}